\documentclass[12pt]{article}
\usepackage{amsmath}
\usepackage{amssymb}
\usepackage{amsfonts}
\newcommand{\ba}{\begin{array}{l}}
\newcommand{\ea}{\end{array}}
\newcommand{\beq}{\begin{equation}}
\newcommand{\eeq}{\end{equation}}
\newcommand{\bea}{\begin{eqnarray}}
\newcommand{\eea}{\end{eqnarray}}
\usepackage{epsfig}
\usepackage{graphicx}
\usepackage{color}

\definecolor{dyellow}{rgb}{1.,0.8,.0}
\definecolor{myblue}{rgb}{.1,.1,.7}
\definecolor{dcyan}{rgb}{.0,.6,.6}
\definecolor{dmagenta}{rgb}{0.6,0.0,0.6}
\definecolor{brown}{rgb}{0.6,0.2,0.}
\definecolor{darkblue}{rgb}{.0,.0,0.5}
\definecolor{darkred}{rgb}{0.75,0.0,0.0}
\definecolor{orange}{rgb}{1.,.6,.0}
\definecolor{dorange}{rgb}{0.8,.4,.0}
\definecolor{darkgreen}{rgb}{0.0,0.6,0.0}
\definecolor{purple}{rgb}{.4,.0,.4}

\def\bc{\begin{center}}
\def\ec{\end{center}}
\def\be{\begin{eqnarray}}
\def\ee{\end{eqnarray}}

\newcommand{\omits}[1]{}

\begin{document}
\begin{center}
{\Large \bf { Modified Friedmann model in Randers-Finsler space of approximate Berwald type as a possible alternative to dark energy hypothesis }}\\
  \vspace*{1cm}
Zhe Chang \footnote{changz@mail.ihep.ac.cn} and  Xin
Li \footnote{lixin@mail.ihep.ac.cn}\\
\vspace*{0.2cm} {\sl Institute of High Energy Physics\\
Chinese Academy of Sciences\\
P. O. Box 918(4), 100049 Beijing, China}\\

\bigskip

\end{center}
\vspace*{2.5cm}

%
\begin{abstract}\baselineskip=30pt
Gravitational field equations in Randers-Finsler space of
approximate Berwald type are investigated. A modified Friedmann
model is proposed. It is showed that the accelerated expanding
universe is guaranteed by a constrained Randers-Finsler structure
without invoking dark energy. The geodesic in Randers-Finsler space
is studied. The additional term in the geodesic equation acts as
repulsive force against the gravity.
 \vspace{1cm}
\begin{flushleft}
\end{flushleft}
\end{abstract}

\newpage
\baselineskip=30pt

Einstein's general relativity connects the Riemann geometry to
gravitation. It is the standard model of gravity. However, up to
now, general relativity still faces problems. One of them is that
the flat rotation curves of spiral galaxies violate the prediction
of Einstein's gravity. Another is related with recent astronomical
observations\cite{Riess}. Our universe is acceleratedly expanding.
This result can not be obtained directly from Einstein's gravity and
his cosmological principle.

The most widely adopted way to resolve these difficulties is the
dark matter and dark energy hypothesis. However, up to now, such
things can not be detected directly from observations. This
situation causes that some physicists imagine the dark matter and
dark energy hypothesis possesses some properties of the ether
hypothesis at the early 20 century. It is reasonable to test the
connection between gravitation and new geometry. Modified Einstein's
gravity may throw new light to the above problems. Models have been
built for alternative to the dark matter hypothesis. The famous one
is the modified Newtonian dynamics\cite{Milgrom}. Models have also
been built for alternative to the dark energy
hypothesis\cite{Bludman}.

Finsler geometry, which takes Riemann geometry as its special case,
is a good candidate to solve the problems mentioned above. In our
previous paper\cite{Finsler DM}, a modified Newton's gravity was
obtained as the weak field approximation of the Einstein's equation
in Finsler space of Berwald type. We have shown that the prediction
of the modified Newton's gravity is in good agreement with the
rotation curves of spiral galaxies without invoking of dark matter
hypothesis. In this Letter, we propose a modified Friedmann model in
Randers-Finsler space of approximate Berwald type for possible
alternative to the dark energy hypothesis.

It is well known that the violation of Lorentz symmetry is one of
the origins of new physics beyond Standard Model. An interesting
case of Lorentz violation, which was proposed by Cohen and
Glashow\cite{Cohen}, is the model of Very Special Relativity (VSR)
characterized by a reduced symmetry SIM(2). In fact, Gibbons, Gomis
and Pope\cite{Gibbons} showed that the Finslerian line element
$ds=(\eta_{\mu\nu} dx^\mu dx^\nu)^{(1-b)/2}(n_\rho dx^\rho)^b$ is
invariant under the transformations of the group DISIM$_b(2)$.
Further investigation of the VSR in Finsler cosmology was
presented\cite{Kouretsis}. In reference\cite{RF}, we have used the
similar method of Gibbons {\em et al.} to study the Lorentz
violations within the framework of Finsler geometry.

Randers space, as a special kind of Finsler space, was first
proposed by G. Randers\cite{Randers}. Within the framework of
Randers space, modified dispersion relation has been
discussed\cite{RF}. A generalized Friedmann-Robertson-Walker (FRW)
cosmology of Randers-Finsler geometry has been also
suggested\cite{Stavrinos}.

The gravity in Finsler space has been studied for a long
time\cite{Takano,Ikeda,Tavakol, Bogoslovsky1}. The gravitational
field equations derived from Riemannian osculating metric were
presented in \cite{Asanov}. The generalized FRW cosmology and the
anisotropies of the universe have been investigated for such a
metric\cite{Kouretsis, Stavrinos}. However, their gravitational
field equations are not consistent with the Bianchi identity and
general covariance principle of Einstein. The gravitational field
equations in Berwald-Finsler space has been written down
explicitly\cite{Lixin}(the Greek indices belong to \{0, 1, 2, 3\}
and the Latin ones to \{1, 2, 3\}), \be\label{field equation of
Berwald}
\left[Ric_{\mu\nu}-\frac{1}{2}g_{\mu\nu}S\right]+\left\{\frac{1}{2}
B^{~\alpha}_{\alpha~\mu\nu}+B^{~\alpha}_{\mu~\nu\alpha}\right\}=8\pi
G T_{\mu\nu}. \ee Berwald space is just a bit more general than the
Riemannian space. Given a Berwald space, all its tangent spaces are
linearly isometric to a common Minkowski space\cite{Ichijyo}. This
property of Berwald space is compatible with the general covariance
principle.

Before dealing with the gravitational field equations, first of all,
we introduce some basic notations of the Finsler geometry\cite{Book
by Bao}. Denote by $T_xM$ the tangent space at $x\in M$, and by $TM$
the tangent bundle of $M$. Each element of $TM$ has the form $(x,
y)$, where $x\in M$ and $y\in T_xM$. The natural projection $\pi :
TM\rightarrow M$ is given by $\pi(x, y)\equiv x$. A Finsler
structure of $M$ is a function\be F :
TM\rightarrow[0,\infty)\nonumber \ee with the following
properties:\\
(i) Regularity: F is $C^\infty$ on the entire slit tangent bundle
$TM\backslash0$.\\
(ii) Positive homogeneity : $F(x, \lambda y)=\lambda F(x,
y)$ for all $\lambda>0$.\\
(iii) Strong convexity: The $n\times n$ Hessian matrix\be
g_{\mu\nu}\equiv\frac{\partial}{\partial
y^\mu}\frac{\partial}{\partial
y^\nu}\left(\frac{1}{2}F^2\right)\nonumber \ee is positive-definite
at every point of $TM\backslash0$.

Throughout the Letter, the lowering and raising of indices are
carried out by the fundamental tensor $g_{\mu\nu}$ defined above,
and its inverse $g^{\mu\nu}$.

In Finsler manifold, there exists a unique linear connection~-~the
Chern connection\cite{Chern}. It is torsion freeness and
metric-compatibility,
 \be
 \Gamma^{\alpha}_{\mu\nu}=\gamma^{\alpha}_{\mu\nu}-g^{\alpha\lambda}\left(A_{\lambda\mu\beta}\frac{N^\beta_\nu}{F}-A_{\mu\nu\beta}\frac{N^\beta_\lambda}{F}+A_{\nu\lambda\beta}\frac{N^\beta_\mu}{F}\right),
 \ee
 where $\gamma^{\alpha}_{\mu\nu}$ is the formal Christoffel symbols of the
second kind with the same form of Riemannian connection, $N^\mu_\nu$
is defined as
$N^\mu_\nu\equiv\gamma^\mu_{\nu\alpha}y^\alpha-A^\mu_{\nu\lambda}\gamma^\lambda_{\alpha\beta}y^\alpha
y^\beta$
 and $A_{\lambda\mu\nu}\equiv\frac{F}{4}\frac{\partial}{\partial y^\lambda}\frac{\partial}{\partial y^\mu}\frac{\partial}{\partial y^\nu}(F^2)$ is the
 Cartan tensor (regarded as a measurement of deviation from the Riemannian Manifold).

The Randers metric is a Finsler structure $F$ on $TM$ with the form
\be\label{Randers metric} F(x,y)\equiv\alpha(x,y)+\beta(x,y) ~,\ee
where \be \alpha(x,y)&\equiv&\sqrt{\tilde{a}_{\mu\nu}(x)y^\mu
y^\nu}\nonumber\\
\beta(x,y)&\equiv&\tilde{b}_\mu(x)y^\mu.\ee Here $\tilde{\alpha}$ is
a Riemannian metric on the manifold $M$. In this Letter, the indices
decorated with a tilde are lowered and raised by
$\tilde{\alpha}_{\mu\nu}$ and its inverse matrix
$\tilde{\alpha}^{\mu\nu}$. A Finsler structure $F$ is said to be of
Berwald type if the Chern connection coefficients
$\Gamma^{\alpha}_{\mu\nu}$ in natural coordinates have no $y$
dependence. Given a Randers space of Berwald type,
Kikuchi\cite{Kikuchi} proved that\be\label{condition of RB}
\tilde{b}_{\mu|\nu}\equiv\frac{\partial\tilde{b}_\mu}{\partial
x^\nu}-\tilde{b}_\kappa\tilde{\gamma}^\kappa_{\mu\nu}=0, \ee where
$\tilde{\gamma}^\kappa_{\mu\nu}$ is the Christoffel symbols of
Riemannian metric $\tilde{\alpha}_{\mu\nu}$. In Randers space of
Berwald type, after some tedious calculations one obtains that \be
\Gamma^\kappa_{\mu\nu}=\tilde{\gamma}^\kappa_{\mu\nu}. \ee The
curvature of Finsler space of Berwlad type is given as \be
R^{~\lambda}_{\kappa~\mu\nu}&=&\frac{\partial
\Gamma^\lambda_{\kappa\nu}}{\partial x^\mu}-\frac{\partial
\Gamma^\lambda_{\kappa\mu}}{\partial
x^\nu}+\Gamma^\lambda_{\alpha\mu}\Gamma^\alpha_{\kappa\nu}-\Gamma^\lambda_{\alpha\nu}\Gamma^\alpha_{\kappa\mu}.\ee
Thus, the curvature of Randers space of Berwald type can be
simplified as \be R^{~\lambda}_{\kappa~\mu\nu}&=&\frac{\partial
\tilde{\gamma}^\lambda_{\kappa\nu}}{\partial x^\mu}-\frac{\partial
\tilde{\gamma}^\lambda_{\kappa\mu}}{\partial
x^\nu}+\tilde{\gamma}^\lambda_{\alpha\mu}\tilde{\gamma}^\alpha_{\kappa\nu}-\tilde{\gamma}^\lambda_{\alpha\nu}\tilde{\gamma}^\alpha_{\kappa\mu}.\ee
This curvature is none other than the curvature of $\tilde{\alpha}$.
The Ricci tensor on Finsler manifold was first introduced by
Akbar-Zadeh\cite{Akbar}. In Finsler space of Berwald type, it
reduces to \be
Ric_{\mu\nu}=\frac{1}{2}(R^{~\alpha}_{\mu~\alpha\nu}+R^{~\alpha}_{\nu~\alpha\mu}).\ee
It is manifestly symmetric and covariant. Apparently the Ricci
tensor will reduce to the Riemann-Ricci tensor if the Cartan tensor
vanish identically. The trace of the Ricci tensor gives the scalar
curvature $S\equiv g^{\mu\nu}Ric_{\mu\nu}$. In order to investigate
the FRW cosmology, we set the Riemannian metric $\tilde{\alpha}$ to
be the Robertson-Walker one \be \tilde{a}_{\mu\nu}={\rm
diag}\left(1, -\frac{R^2(t)}{1-kr^2}, -R^2(t)r^2,
-R^2(t)r^2\sin^2\theta\right),\ee  where $k=0,\pm1$ for a flat,
closed and hyperbolic geometry respectively. Unfortunately, such a
Randers space of Berwald type is just the Riemannian space. That is
the condition (\ref{condition of RB}) only has solution
$\tilde{b}=0$. Here, we set $\tilde{b}_\mu=(\tilde{b}_0,0,0,0)$ for
satisfying the requirement that the universe is homogenous and
isotropic. If $\tilde{b}_0$ is sufficient small, the space can be
regarded as a Berwald space approximately. On such approximation, we
just neglect the term proportion to $\int
\frac{\partial\Gamma}{\partial y} dx$ in the field equation
(\ref{field equation of Berwald}).

After some tedious but straightforward calculations, we obtain
following nonzero components of curvature in Randers space of
approximate Berwald type \be
Ric_{00}&=&-3\frac{\ddot{R}}{R}\tilde{a}_{00},\nonumber\\
Ric_{ij}&=&-\left(\frac{\ddot{R}}{R}+2\frac{\dot{R}^2}{R^2}+\frac{2k}{R^2}\right)\tilde{a}_{ij},\nonumber\\
S&=&-6\frac{\alpha}{F}\left(\frac{\ddot{R}}{R}+\frac{\dot{R}^2}{R^2}+\frac{k}{R^2}\right)\nonumber\\
&&-3\frac{\ddot{R}}{R}\tilde{a}_{00}\frac{\alpha^2}{F^2}\left(\frac{\beta}{F}\tilde{a}_{00}\frac{y^0}{\alpha}\frac{y^0}{\alpha}-2\tilde{a}_{00}\frac{y^0}{\alpha}\tilde{b}^0\right)\nonumber\\
&&-3\left(\frac{\ddot{R}}{R}+2\frac{\dot{R}^2}{R^2}+\frac{2k}{R^2}\right)\tilde{a}_{ij}\frac{\alpha^2}{F^2}\left(\frac{\beta}{F}\tilde{a}_{ij}\frac{y^i}{\alpha}\frac{y^j}{\alpha}\right).
\ee The terms $B^{~\alpha}_{\alpha~\mu\nu}$ and
$B^{~\alpha}_{\mu~\nu\alpha}$ vanish in Randers space of approximate
Berwald type, where \be
B_{\mu\nu\alpha\beta}=-A_{\mu\nu\lambda}R^{~\lambda}_{\theta~\alpha\beta}y^\theta/F.\ee
In the left side of the field equations, only symmetric part is
left. Thus, we should set the energy-momenta tensor as\be
T^\mu_\nu={\rm diag}(\rho, -p,-p,-p),\ee where $\rho\equiv\rho(x)$
and $p\equiv p(x)$ is the the energy density and pressure of the
cosmic fluid respectively. The $0-0$ component of the field
equations (\ref{field equation of Berwald}) gives the modified
Friedmann equation \be\label{MFE}
\frac{\alpha}{F}\left(\frac{\dot{R}^2}{R^2}+\frac{k}{R^2}\right)-\frac{1}{2}\frac{\ddot{R}}{R}\tilde{a}_{00}\frac{\alpha^2}{F^2}\left(\frac{\beta}{F}\tilde{a}_{00}\frac{y^0}{\alpha}\frac{y^0}{\alpha}-2\tilde{a}_{00}\frac{y^0}{\alpha}\tilde{b}^0\right)\nonumber\hspace{5cm}\\
+\frac{1}{2}\left(\frac{\ddot{R}}{R}+2\frac{\dot{R}^2}{R^2}+\frac{2k}{R^2}\right)\tilde{a}_{ij}\frac{\alpha^2}{F^2}\left(\frac{\beta}{F}\tilde{a}_{ij}\frac{y^i}{\alpha}\frac{y^j}{\alpha}\right)=\frac{8\pi
G}{3}\rho. \ee By making use of the modified Friedmann equation
(\ref{MFE}) and omitting the $O(b^2)$ term, we obtain the $i-i$
component of the field equations (\ref{field equation of Berwald})
 \be\label{equation
of R}
\frac{\alpha}{F}\frac{\ddot{R}}{R}\left(1+\frac{\alpha}{F}\left(\frac{\beta}{F}\tilde{a}_{00}\frac{y^0}{\alpha}\frac{y^0}{\alpha}-2\tilde{a}_{00}\frac{y^0}{\alpha}\tilde{b}^0\right)\right)=-\frac{4\pi
G}{3}(\rho+3p).\ee From the equation (\ref{equation of R}), one can
see clearly that the accelerated expanding universe ($\ddot{R}>0$)
is guaranteed by the constraint \be\label{inequality}
1+\frac{\alpha}{F}\left(\frac{\beta}{F}\tilde{a}_{00}\frac{y^0}{\alpha}\frac{y^0}{\alpha}-2\tilde{a}_{00}\frac{y^0}{\alpha}\tilde{b}^0\right)<0
,\ee while the energy density and pressure of the cosmic fluid keep
positive. Since the Finsler structure $F$ and Riemannian length
element $\alpha$ are positive, a direct deduction from
(\ref{inequality}) is  \be\label{inequality1}
\tilde{b}_0&<&-\frac{1}{\frac{y^0}{\alpha}((\frac{y^0}{\alpha})^2-2)},\\
\label{inequality2}\frac{y^0}{\alpha}&>&\sqrt{2}.\ee The positive
Finsler structure $F$ gives that $\tilde{b}_0\tilde{b}^0<1$. So that
the complete constraint on Randers-Finsler structure to support
accelerated expanding universe is
 \be
 -1<\tilde{b}_0<-\frac{1}{\frac{y^0}{\alpha}((\frac{y^0}{\alpha})^2-2)}.
 \ee
It means that a negative $\tilde{b}_0$ provides an effective
repulsive force in the course of universe expanding.

This fact also can be observed clearly from the geodesic with
constant Riemanian speed. Following the calculus of variations, one
get the geodesic equation of Finsler space\cite{Book by Bao}
\be\label{geodesic of F}
\frac{d^2\sigma^\lambda}{d\tau^2}+\gamma^\lambda_{\mu\nu}\frac{d\sigma^\mu}{d\tau}\frac{d\sigma^\nu}{d\tau}=\frac{d\sigma^\mu}{d\tau}\frac{d}{d\tau}\left(\log
F(\sigma, \frac{d\sigma}{d\tau})\right). \ee Deducing from
(\ref{geodesic of F}), we obtain the geodesic of Randers space with
constant Riemanian speed (namely, $\alpha(\frac{d\sigma}{d\tau})$ is
constant) \be\label{geodesic of R}
\frac{d^2\sigma^\lambda}{d\tau^2}+\tilde{\gamma}^\lambda_{\mu\nu}\frac{d\sigma^\mu}{d\tau}\frac{d\sigma^\nu}{d\tau}+\tilde{a}^{\lambda\mu}f_{\mu\nu}\alpha\left(\frac{d\sigma}{d\tau}\right)\frac{d\sigma^\nu}{d\tau}=0,
\ee where $f_{\mu\nu}\equiv\frac{\partial\tilde{b}_\mu}{\partial
x^\nu}-\frac{\partial\tilde{b}_\nu}{\partial x^\mu}$. The geodesic
equation of Randers space (\ref{geodesic of R}) has clearly physical
meaning. The last term, which is proportional to the asymmetrical
term $f_{\mu\nu}$, acts as electromagnetic force. The term
$\tilde{b}_\mu$ can be regarded as the electromagnetic potential.
The negative $\tilde{b}_0$ means that the``electromagnetic" force
$f$ is repulsive and against the popular attractive force.

Since the Finsler structure depends on both coordinates and
velocities, it is important to investigate the physical meaning of
the velocity dependence.
The term $\frac{y^0}{\alpha}$\cite{RF} involved in
(\ref{inequality1}) represents the energy--to--mass ratio. The upper
bound of the dimensionless parameter $\tilde{b}_0$ gives a criteria
that the repulsive effect equal to the attractive one. It means that
the universe is expanding with constant speed while $\tilde{b}_0$
equal to its upper bound. The particle have enough energy to fight
against the attractive force while $\tilde{b}_0$ satisfies the
constraint (\ref{inequality1}).
\bigskip

\centerline{\large\bf Acknowledgements} \vspace{0.5cm}
 We would like to thank Prof. H. Y. Guo and C. G. Huang for useful discussions. The
work was supported by the NSF of China under Grant No. 10575106 and
10875129.


\begin{thebibliography}{999}
\bibitem{Riess}A. G. Riess, {\it et al}.,
Astrophys J. {\bf 117}, 707 (1999); S. Perlmutter, {\it et al}.,
Astrophys J. {\bf 517}, 565 (1999); C. L. Bennett, {\it et al}.,
Astrophys J. {\bf 148} (Suppl), 1 (2003).
\bibitem{Milgrom}M. Milgrom, Astrophys. J. {\bf 270}, 365 (1983).
\bibitem{Bludman}S. Bludman, arXiv:astro-ph/0605198.
\bibitem{Finsler DM}Z. Chang and X. Li, Phys. Lett. B {\bf
668}, 453 (2008).
\bibitem{Cohen}A. G. Cohen and S. L. Glashow, Phys. Rev. Lett. {\bf 97},
021601 (2006).
\bibitem{Gibbons}G. W. Gibbons, J. Gomis and C. N. Pope, Phys. Rev. D {\bf 76}, 081701
(2007).
\bibitem{Kouretsis}A. P. Kouretsis, M. Stathakopoulos and
P. C. Stavrinos, arXiv:gr-qc/0810.3267.
\bibitem{RF}Z. Chang and X. Li, Phys. Lett. B {\bf 663}, 103 (2008).
\bibitem{Randers}G. Randers, Phys. Rev. {\bf 59}, 195 (1941).
\bibitem{Stavrinos}P. C. Stavrinos, A. P. Kouretsis and
M. Stathakopoulos, arXiv:gr-qc/0612157.
\bibitem{Takano}Y. Takano, Lett. Nuovo Cimento {\bf 10}, 747
(1974).
\bibitem{Ikeda}S. Ikeda, Ann. der Phys. {\bf 44}, 558 (1987).
\bibitem{Tavakol}R. Tavakol and N. van den Bergh, Phys. Lett. A {\bf 112}, 23
(1985).
\bibitem{Bogoslovsky1}G. Yu. Bogoslovsky, Phys. Part. Nucl. {\bf 24}, 354
(1993).
\bibitem{Asanov}G.S.Asanov, {\it Finsler Geometry, Relativity and Gauge Theories}, Reidel
Pub.Com., Dordrecht, 1985.
\bibitem{Lixin}X. Li and Z. Chang, arXiv: gr-qc/0711.1934.
\bibitem{Ichijyo}Y. Ichijy\={o}, {\it Finsler manifolds modeled on a Minkowski
space}, J. Math. Kyoto Univ. {\bf 16-3}, 639 (1976).
\bibitem{Book by Bao}D. Bao, S. S. Chern and Z. Shen, {\it An
Introduction to Riemann--Finsler Geometry}, Graduate Texts in
Mathmatics {\bf 200}, Springer, New York, 2000.
\bibitem{Chern}S. S. Chern, Sci. Rep. Nat. Tsing Hua Univ. Ser. A
{\bf 5}, 95 (1948); or Selected Papers, vol. II, 194, Springer 1989.
\bibitem{Kikuchi}S. Kikuchi, Tensor, N.S. {\bf 33}, 242 (1979).
\bibitem{Akbar}H. Akbar-Zadeh, Acad. Roy. Belg. Bull. Cl. Sci. (5)
{\bf 74}, 281 (1988).

\end{thebibliography}
\end{document}